\documentstyle[12pt]{article}
\begin{document}
\title{ High order amplitude equation for steps on creep curve}
\author{Mulugeta Bekele$^{1,}$\thanks {On leave from Department of Physics, 
Addis Ababa University, Addis Ababa, Ethiopia} and G. Ananthakrishna$^2$\\
$^1$ Department of Physics, $^2$ Materials Research Centre\\
Indian Institute of Science, Bangalore 560 012, India\\}
\date{}
\maketitle
\begin{abstract}
We consider a model proposed by one of the authors for a type of
plastic instability 
found in creep experiments which reproduces a number of experimentally 
observed features. The model consists of three coupled non-linear 
differential equations describing the evolution of three types of 
dislocations. The transition to the instability
has been shown to be via Hopf bifurcation leading to limit cycle 
solutions with respect to physically relevant drive parameters.
Here we use reductive perturbative method to extract an amplitude  
equation of up to {\it seventh} order to 
obtain an approximate analytic expression for the order parameter. 
The analysis also enables us to obtain the bifurcation (phase) diagram 
of the instability. We find that while supercritical bifurcation  
dominates the major part of the instability region, subcritical 
bifurcation gradually takes over at  {\it one end} of the region.
These results are compared with the known experimental results.
Approximate analytic expressions for the limit cycles 
for  different types of bifurcations are shown to agree 
with their corresponding  numerical solutions of the equations 
describing the model. The analysis also shows that high order 
nonlinearities are important in the problem. This approach further 
allows us to map the theoretical parameters to the experimentally 
observed macroscopic quantities.
\end{abstract}

\newpage
\section{Introduction}
Instabilities in plastic flow has been an object of attention for a
long time in metallurgical literature. Experimentally, there are 
basically three modes of deformation of a specimen. The best known 
and mostly widely studied form of the instability arises when the specimen 
is subjected to a constant rate of tensile deformation commonly referred to as
constant  strain rate test \cite{brind,bod}. Clearly, this method 
of deformation is conceptually difficult to understand 
since the specimen is subjected to  a predetermined response ( i.e., a 
constant rate of deformation), and the force or the stress developed in the
sample is sought to be measured. Under normal conditions, one finds a smooth 
stress-strain curve. However, when the system is 
in the regime of instability (i.e.,  for some values of the material parameters),
the stress-strain curve exhibits repeated load drops.
Each of the load drops is associated with the formation and propagation of
dislocation bands \cite{chih}. 
There is another form of testing, where, the deformation is carried out
keeping the stress rate  fixed. Again, under normal conditions one 
finds a smooth stress-strain curve. However, when the material is 
deformed in the instability regime of the parameter space, 
one finds a stepped response in the stress-strain curve.
This method of deformation is equally popular among experimentalists 
for the study of the instability. 
However, conceptually the simplest  form 
of the instability \cite{hall} manifests when the material is subjected
to  a creep test wherein a force is applied and the response in the
form of elongation of the specimen is measured. Here again, under normal
conditions, the strain-time
curve is smooth.  Under certain metallurgical conditions, one sees steps
on creep curve suggesting a form of instability \cite{hall,dasi}.
It is in the former two types of testing, that the plastic instability  
manifests 
much more easily than in the last case and hence these two modes of 
deformation
are usually adopted. In contrast, the phenomenon of steps on 
creep curve, which is the subject of the present discussion, is seen 
in much fewer instances \cite{dasi,ste}.
Instabilities occurring in all these forms are considered to be of common 
origin. The phenomenon is referred to as
the Portevin-Le Chatelier (PLC) effect or the jerky flow and is seen
in several metals such as commercial aluminium, brass, alloys of aluminium 
and magnesium \cite{brind}. In the case of constant strain rate case, 
it is observed only in a window of strain rates and 
temperature.

It is generally agreed that the 
microscopic origin of the instabilities arises due to the interaction of 
dislocations with mobile point defects and is
referred to as dynamic strain ageing. This leads to negative strain rate 
characteristic of the flow stress. The basic idea was formulated by
Cottrell \cite{cott} few decades ago. 
Early phenomenological models including Cottrell's theory and its extensions
\cite{cott,mcor} do not deal with time development. In contrast, techniques 
of dynamical systems addresses precisely this aspect. Recently, there 
has been a resurgence of interest in plastic instabilities 
[9-13] in light of introduction of new methodology 
borrowed from theory of dynamical systems. This has helped to obtain new 
insights hitherto not possible [10-18]. One of the aims 
of such theories is to be able to relate the microscopic dislocation 
mechanisms to the measurable macroscopic quantities.

An attempt to understand the problem
in the above perspective was made by Ananthakrishna and coworkers 
several years ago \cite{gads2,gava}. The basic idea was to describe 
the problem 
from the point of view of far-from-equilibrium transition,
wherein the new temporal order could be described as a cooperative
phenomenon \cite{nico,hak}. In a series of papers 
\cite{gads1,gads3}, starting from an extended
Fokker-Planck equation for the velocity of dislocation segments,  
these authors 
arrived at a model which consisted of three types of dislocations
and  some transformations between them \cite{gads2}. The basic 
idea could  be  summarized by  stating that  limit  cycle
solutions arise  due  to   nonlinear interaction between  three
different types of dislocations, suggesting a new mathematical
mechanism for  the instability.  Even though the  spatial
inhomogeneous structure was ignored and only the temporal oscillatory
state was sought to be described, the model and its extensions to the
case of constant strain rate test \cite{gava}, proved to be very 
successful in that
it could explain most of the experimentally observed features 
such as the existence of bounds on strain rate for the PLC effect to
occur, the negative strain rate sensitivity, etc. 
\cite{bod,cott,gava}. One other important prediction which is a 
direct consequence of   the dynamical basis of the model is 
the existence of chaotic stress drops in a range of strain rates 
\cite{gava1,gaj}. 
{\it Recently, there has been several attempts which verifies this 
prediction} [17-20]. {\it Indeed this verification 
suggests that these few modes represent the collective degrees of freedom 
of dislocations}. (Note that the spatially
extended nature of system implies infinite degrees of freedom.) 
From this point of view, dealing with the temporal aspect appears to be
justified. Description of  the phenomenon  which
includes  the initiation  and propagation of the bands during the PLC
effect has also been recently attempted \cite{gscrp1}.

Since the introduction of bifurcation theory into this field several
years ago by our group \cite{gads2,gava}, several other groups 
have also undertaken similar lines of attack [9-12],
\cite{aifa}. In the process, we feel that finer aspects of dynamical 
systems have been glossed over in this field. For
instance, one often finds that casual remarks are made about fast and
slow modes without actually going through the procedure of
demonstrating the existence of such modes and eliminating the
fast modes in favour of the slow ones \cite{aifa}. In addition, under the
adiabatic elimination, the resulting modes which serve as order
parameter variables are very complicated functions of the original
modes. Yet, hand waiving arguments have been used in building models
which we believe are technically suspect.

In our recent work \cite{mbga2} we showed how  under certain conditions,
one of the variables of the model could be adiabatically eliminated since
the time constant of this mode can be chosen to be much faster than the 
other two ( i.e., for low values of a parameter  $b_0$, see below). 
We then derived the equation for the order parameter for the reduced model. 
We found both supercritical and subcritical bifurcation  within the range of 
applicability. We also found that the results were in good agreement both 
with the reported experimental results and with the numerical solution of 
the model.  However, eliminating one of the variables entirely restricts the 
applicability of the analysis to the two-dimensional plane of the parameter 
space (parameters $a$ and $c$, see below). In addition, we also found that 
even within the limited domain, very high order nonlinearities control most 
of the bifurcation domain. 

The purpose of the present work is to perform the analysis by keeping all 
the three modes in the model and to explore the entire instability domain
spanned by all the three parameters ($a$, $b_0$ and $c$). In addition, this 
analysis should help us to verify if high order nonlinearities could control 
part of the subcritical bifurcation as found in our recent analysis 
\cite{mbga2}.  This will  help us to investigate the full nature of the 
bifurcation in detail. We use reductive perturbative method and extract the 
complex order parameter  which is directly related to the amplitude and the 
frequency of the jumps on the creep curve. The analysis should also help us 
to compare the results with the experimental ones. The expression for the 
order parameter is checked by comparing it with the numerical solution of 
the model. 
   
In what follows (section 2) we present a brief summary of the model.
In section 3 we use the reductive perturbative method to extract the 
amplitude equation up to {\it seventh} (septic) order. This enables 
us not only to determine the nature of bifurcation (i.e. supercritical or 
subcritical) exhibited by the model but also gives us an expression for the 
order parameter over most of the instability domain. In section 4,  the 
approximate limit cycle solution obtained through the amplitude equation is 
compared with the experimental results as well as with the numerical 
solution of the model. Section 5 contains summary and discusses our results.

\section{ A Model for Steps on Creep Curve}

We start with a brief summary of the model. The details of the model
can be found in the original references \cite{gads2}.  
The model consists  of the
mobile dislocations, the immobile dislocations  and another type
which mimics the Cottrell's type, which are dislocations with clouds of
solute atoms. Let the corresponding densities be $N_m$, $N_{im}$ and
$N_i$, respectively.  The rate equations for the densities of
dislocations are:
\begin{eqnarray}
\frac{dN_m}{dt} & = & \theta V_m N_m - \beta N_m^2 -\beta N_m N_{im} +\gamma N_{im}
         -\alpha_m N_m\,,
\\
\frac{dN_{im}}{dt} & = &\beta N^2_m -\beta N_{im} N_m -\gamma N_{im} +\alpha_i N_i,
\\
 \frac{dN_i}{dt} & = & \alpha_m N_m - \alpha_i N_i.
\end{eqnarray}

\noindent
The first term in Eq.(1) is
the rate of production of dislocations due to cross glide with a rate
constant $\theta V_m$ , where $V_m$ is the velocity of the mobile 
dislocations
which in general depends on some power of the applied stress,
$\sigma_a$. The second term refers to two mobile dislocations either
annihilating or immobilizing. The third term
also represents the annihilation of a mobile dislocation with an
immobile one.  The fourth term represents the remobilization of the
immobile dislocations due to stress or thermal activation ( see $\gamma
N_{im}$ in Eq. 2). The last term represents the immobilization of
mobile dislocations either due to solute atoms or due to other pinning
centers.   $\alpha_m$ refers to the concentration of the solute atoms
which participate in slowing down the mobile dislocations.  Once a mobile
dislocation starts acquiring  solute atoms we regard it as a new type of
dislocation, namely the Cottrell's type $ N_i$. This process is
represented as an incoming term in Eq.(3). As they acquire more and
more solute atoms they will slow down and eventually stop
the dislocation entirely. At this point, they are considered to have
transformed to $N_{im}$. This process is represented by the loss term
in Eq.(3) and a gain term in Eq.(2).  These equations can be cast into
a dimensionless form by using scaled variables:
\begin{equation}
x = N_m (\frac{\beta}{\gamma}),
y = N_{im}(\frac{\beta}{\theta V_m}),
z = N_i (\frac{\beta\alpha_i}{\gamma\alpha_m}) ,\tau=\theta V_m t,
\end{equation}
to get
\begin{eqnarray}
\dot{x} & = & (1-a)x -b_0x^2 -xy +y,
\\
\dot{y} & = & b_0\left(b_0x^2 -xy-y+az\right),
\\
\dot{z} & = & c(x-z),
\end{eqnarray}
\noindent
where the dot represents differentiation with respect to $\tau$ while
$a =  \frac{\alpha_m}{\theta V_m},
b_0 = \frac{\gamma}{\theta V_m}, {\rm and }\, c = \frac{\alpha_i}{\theta V_m}$.
Eqs.(5-7) are coupled set of nonlinear equations which support limit
cycle solutions for a range of parameters $a,b_0$  and $ c$, that 
are physically relevant. $a$ refers to the concentration of
the solute atoms, $b_0$ refers to the reactivation of immobile
dislocations and $c$ to the time scales over which the slowing down
occurs. The dependence  on stress and temperature appears through $V_m$.
We have demonstrated the existence of limit cycle solutions and 
also obtained approximate closed form solutions for the limit cycles 
\cite{gads2}. In addition, the model has been studied numerically.
Using the Orowan equation which relates the rate of change
of strain($\dot{S}$) to dislocation density and the mean velocity:
$\dot{S}=bN_mV_m$,  with $b$ as the
Burger's vector, steps on the creep curve follow automatically since
the densities of dislocations are oscillatory.  Several experimental
results are reproduced \cite{gads2}.

\section{ Reductive Perturbative Approach}

We briefly outline the reductive perturbative approach to problems of 
formation of new states of order in far-from-equilibrium situations. 
Transitions occurring in these systems are quite analogous to equilibrium 
phase transitions. The general idea is to construct a `potential like 
function' for the `order parameter' like variable in the neighbourhood of 
the critical value of the drive parameter. This would permit the use of the 
methods developed in equilibrium phase transitions for further analysis.
 
Near the point of Hopf bifurcation of the system (Eqs.5-7), corresponding 
to a value near the critical drive parameter, a pair of complex conjugate
eigenvalues and another real negative eigenvalue exist for the
linearized system of equations around the steady state. As we approach 
the critical value from the stable side, the real part of the pair of
complex conjugate eigenvalues approaches  zero from the negative side and 
hence the corresponding eigen directions have slow time scale. As we enter 
the instability region, these real parts become positive. In contrast, the 
effect of the change in the drive parameter on the real negative eigenvalue 
is negligible.  Thus, {\it while the two eigenvectors corresponding to
the pair of complex conjugate eigenvalues are slow modes the eigenvector 
corresponding to the real negative eigenvalue is a fast (and decaying) mode}. 
For this reason, the slow modes determine the formation of new states of 
order.  {\it The reductive perturbative method is a method where the slow 
enslaving dynamics is extracted in a systematic way} [27-32].  The method 
involves in first finding the critical eigenvectors corresponding to the 
bifurcation point and expressing the general solution as a linear 
combination of these vectors. The effect of the nonlinearity is handled 
progressively using multiple scales method. The equations governing the 
complex order parameter takes the form of Stuart-Landau equation and 
corresponds to the time-dependent Ginzburg-Landau equation for a 
homogeneous medium. (Henceforth, we will also refer to this equation 
as amplitude equation.) On the other hand, the asymptotic solution, which is 
a limit cycle, collapses to the sub-space spanned by the slow modes with no 
trace of the fast mode. It may be worth emphasizing that this method is 
essentially the same as reduction to center manifold. Indeed, the 
equivalence of the center manifold theory [33-35] with the reductive 
perturbation has been established \cite{chen}. Other techniques of
extracting amplitude equations have been devised whose end results are 
basically the same.  For instance, perturbative renormalization group 
method \cite{golden,chen} and its recent extension on the basis of
envelope theory \cite{kuni} has also been developed as a tool for global 
asymptotic analysis which can be used to extract the amplitude equations.

We start with the Eqs.(5-7). There is only one fixed point defined by: 
\begin{equation}
x_{a} = z_{a} = {{1-2a+[(1-2a)^2+8b_0]^{1/2}}\over{4b_0}},
{\rm and}\,\,,y_{a}= \frac{1}{2}.
\end{equation}

\noindent
Defining new variables which are deviations from the fixed point
\begin{equation}
X=x-x_a,Y=y-y_a, {\rm and}\,\,, Z=z-z_a,
\end{equation}
\noindent
Eqs.(5-7) become 
  \begin{eqnarray}
  \dot{X} & = & -(\alpha X +\chi Y +b_0X^2 +XY ),
  \\
\dot{Y} & = & - b_0\left(\Gamma X+\delta Y-a Z -b_0 X^2 + XY \right),
   \\
\dot{Z} & = & c(X-Z).
\end{eqnarray}
where
\begin{eqnarray}
   \alpha \,=\, a +2b_{0}x_{a} +y_{a} -1,
   \chi \,=\,  x_{a} -1,
   \nonumber
\\
   & &
\\ 
   \Gamma\,=\, y_{a} -2b_{0}x_{a},
   \delta\,=\, x_{a} +1.
   \nonumber
\end{eqnarray}

\noindent
Eqs.(10-12) will be solved reductive perturbatively.
(We note here that it is possible to reduce these system
of equations to only two by adiabatically eliminating Eq. (10). 
This is done by noting that when $b_0$ and $c$ is much smaller than $a$,
 by rescaling Eq.(10), it can be shown to be fast variable and hence 
can be adiabatically eliminated. This was what was done in ref \cite{mbga2}.)
Writing these equations as a matrix equation where the nonlinear part appears
separately from the linear part, we get
\begin{equation}
        {{d\vec{R}}\over{d\tau}}= {\bf L}\vec{R} + \vec{N}
\label{mtrxeq}
\end{equation}
\noindent
where
\begin{equation}
    \vec{R}=\left(
                       \begin{array}{c}
                          X\\
                          Y\\
                          Z
                       \end{array}
                  \right),  
\end{equation}
\begin{equation}
       {\bf L}=\left(
                 \begin{array}{ccc}
                    -\alpha         &-\chi             & 0 \\
                   -b_0\Gamma       &-b_0\delta        & ab_0\\ 
	            c               &0                 &-c
                 \end{array}
                 \right),
      \end{equation}
\noindent
and the nonlinear part, $\vec{N}$, is given by
\begin{equation}
   \vec{N}=\left(
                \begin{array}{c}
                  -b_0X^2 - XY \\
                   b_0(b_0X^2 - XY)\\
                   0 
                \end{array}
                \right). 
\end{equation}

Consider the stability of the fixed point as a function of the 
parameter $c$. The eigenvalues, $\lambda_i$,$i=1,0,-1$, of the matrix 
${\bf L}$ are determined from the cubic equation:
\begin{equation}
\lambda^3 - T \lambda^2 + P \lambda - \Delta = 0,
\end{equation}
\noindent
where
\begin{equation}
   T = -(\alpha + \delta b_0 + c),
\end{equation}
\begin{equation} 
   P = \delta b_0c + \alpha(\delta b_0 + c) - \chi\Gamma b_0,
\end{equation}
\noindent
and
\begin{equation}
\Delta = -b_0c[\alpha\delta + \chi(a - \Gamma)].
\end{equation}
\noindent
The fixed point becomes unstable when one of the following conditions
are violated \cite{murray}:
\begin{equation}
T < 0,  \Delta < 0 \,\, or \,\, \Delta - PT > 0.
\label{ineqs}
\end{equation}
\noindent
It is easy to show that the first two inequalities are not violated.
Substituting for $\Delta$, $P$ and $T$ in the third inequality of 
Eq.(\ref{ineqs}), we get
\begin{equation}
(\alpha + \delta b_0)c^2 + [(\alpha + \delta b_0)^2 - \chi a b_0]c + b_0(\alpha + \delta b_0)(\alpha\delta - \chi\Gamma) < 0
\label{ineq}
\end{equation}
\noindent
as the condition for instability. Using the equality sign in 
Eq.(\ref{ineq}) gives us the critical value 
of the drive parameter, $c=c_0$, for a given values of $a$ and $b_0$.  
For $c < c_0$, the fixed point is unstable. Since $c$ is non-negative
(negative $c$ is unphysical), we get a unique $c_0$ for the allowed pair of 
$a$ and $b_0$ values within the instability.
Fig.1 shows a three-dimensional plot of the instability region involving 
all the three parameters of the model. (Note that since all the variables 
and the parameters are dimensionless quantities, we have all figures plotted 
in dimensionless quantities.)

To get approximate analytical  solution of Eq.(14), we follow the
reductive perturbative approach similar to that used by Mashiyama
{\it et al}. \cite{mashi} and Richter {\it et al.} \cite{rich}. 
We choose 
$c=c_0 (1-\epsilon)$ with $0 < \epsilon \ll 1$ and 
write the matrix ${\bf L}$ as a sum of two matrices, 
${\bf L}= {\bf L}_0 +\epsilon{\bf L}_1$,
where ${\bf L}_0$ is the matrix ${\bf L}$ evaluated for $c=c_0$ and 
\begin{equation}
   {\bf L}_1\equiv\left(
                \begin{array}{ccc}
                     0        & 0           & 0 \\
                     0        & 0           & 0 \\
                    -c_0      & 0           & c_0 
               \end{array}
                \right).
\end{equation}
\noindent
The eigenvalues of ${\bf L}_0$ are,
\begin{equation}
\lambda_{1,-1} = \pm \imath \omega,\,\, and\,\, \lambda_0 = T,
\end{equation}
\noindent
where $\omega^2=P$ ($P$ and $T$ being evaluated at $c=c_0$). 
Taking the solution for $\vec{R}$ as a growth out of
the critical eigenmodes, we express it as a linear combination of
these eigenmodes:
\begin{equation}
\vec{R}(\tau) \,=\, \Psi \, e^{\imath\omega\tau} \vec{r}_{1} \,+\, \Psi_{0}
e^{\lambda_{0} \tau} \vec{r}_{0} \,+\, \Psi^{*} e^{-\imath\omega\tau} \vec{r}_{1}^{*} \,=\,
\sum ^{-1}_{j=1}  \Psi _{j}  e^{\lambda_{j}  \tau}  \vec{r}_{j}
\label{lincomb}
\end{equation}
\noindent
where $\vec{r}_j$'s are right-eigenvectors defined by 
${\bf L}_0\vec{r}_j=\lambda_j\vec{r}_j$ with 
$\vec{r}_{-1}=\vec{r}_1^*$. We also introduce left-eigenvectors, 
$\vec{s}_j^T$, defined by 
$\vec{s}_j^T{\bf L}_0=\lambda_j\vec{s}_j^T$, where $T$ stands for the
transpose.  Substituting this expression for  $\vec{R}$
in the matrix equation, Eq.(\ref{mtrxeq}), and multiplying both sides of 
the equation by one of the left-eigenvectors, we get an equation  
governing the corresponding amplitude:
\begin{equation}
e^{\lambda_j\tau} \,\, \frac {d \Psi_{j}} {d\tau} \,=\, \epsilon
\sum_{k}  \mu_{jk} \Psi_k e^{\lambda_{k} \tau} \,+\, 
\sum_{l,m \,,m \leq l} g_{jlm} \Psi_{l} \Psi_{m}
e^{(\lambda_{l} \,+\, \lambda_m) \tau}.
\label{ampeqn}
\end{equation}
\noindent
Expressions for the coefficients $\mu_{jk}$ and $g_{jlm}$ are given 
in Eq.(C11) and Eqs.(C13,C14), respectively, in Appendix C.

We express $\Psi_j$ as a power series expansion in $\epsilon^{\frac{1}{2}}$:
\begin{equation}
\Psi_{j} \,=\, \epsilon^{\frac {1}{2}} \psi^{(1)}_{j} \,+\, \epsilon
\psi^{(2)}_{j} \,+\,\epsilon^{\frac {3}{2}} \psi^{(3)}_{j} \,+\,  ....,
\end{equation}
\noindent
and introduce multiple time scales such that
\begin{equation}
\frac {d}{d\tau} \,=\, \frac{\partial}{\partial \tau} \,+\, \epsilon
\frac{\partial}{\partial \tau_{1}} \,+\, \epsilon^{2} \frac{\partial}
{\partial \tau_{2}} \,+\, ....,
\end{equation}
\noindent
where $\tau_1=\epsilon\tau$, $\tau_2=  
\epsilon^{2} \tau$,...
Substituting these expressions for $\Psi_{j}$ and $\frac{d}{d\tau}$ in 
the equation for the amplitudes, Eq ($\ref{ampeqn}$) , we successively solve
by equating terms of the same order in powers of $\epsilon$. First, terms of 
$\cal O$($\epsilon^{\frac{1}{2}}$) give
\begin{equation} 
\frac {\partial \psi^{(1)}_{j}}
{\partial \tau}  \,=\,  0,
\end{equation}
\noindent
implying that  $\psi^{(1)}_{j}$ is constant in the time scale of $\tau$.
$\cal O$($\epsilon)$ terms give the equation
\begin{equation}
\frac {\partial \psi^{(2)}_{j}} {\partial \tau}
\,=\, \sum_{k,l,l \leq k} g_{jkl}  \psi^{(1)}_{k}
\psi^{(1)}_{l} e^{(\lambda_{k} \,+\, \lambda_{l} -\lambda_{j}) \tau},
\end{equation}
\noindent
which upon integration gives
\begin{equation}
\psi^{(2)}_{j} e^{\lambda_{j} \tau} \,=\, \sum _{k,l,\, l \leq k} h_{jkl}
\psi^{(1)}_{k} \psi^{(1)}_{l} e^{(\lambda_{k} \,+\, \lambda_{l}) \tau},
\end{equation}
\noindent
where $h_{jkl}=\frac{g_{jkl}}{\lambda_k + \lambda_l -\lambda_j}$.
$\cal O$($\epsilon^{\frac{3}{2}}$) terms give the equation
\begin{equation}
\frac{\partial \psi^{(3)}_{j}}{\partial \tau} \,+\, \frac{\partial
\psi^{(1)}_{j}}{\partial \tau_{1}} \,=\, \sum_{k} 
\mu_{jk} \psi^{(1)}_{k} e^{(\lambda_{k}-\lambda_{j}) \tau} \,+\,
\sum_{k,l, \,,l \leq k} g_{jkl}
(\psi^{(1)}_{k} \psi^{(2)}_{l} \,+\, \psi^{(2)}_{k} \psi^{(1)}_{l})
e^{(\lambda_{k} \,+\,  \lambda_{l} -\lambda_{j}) \tau}.
\end{equation}
\noindent
Using compatibility condition, we match terms that are varying 
on a slow time scale found on both sides of the equality and extract the 
slow dynamics:    
\begin{equation}
\frac {\partial \psi^{(1)}_{j}} {\partial \tau_{1}} \,=\, \mu_{jj}
\psi^{(1)}_{j}\,+\, \eta_{j} \vert \psi^{(1)}_1 \vert^{2}
\psi^{(1)}_{j}.
\end{equation}
\noindent
Expression for $\eta_j$ is given in Eq.(C12) in Appendix C.
Since we are interested in asymptotic solutions, the equations governing 
the oscillatory amplitude $\Psi$ and its complex conjugate $\Psi^*$
are our concern. (The subscript j=1 is left out from 
$\Psi_1$ for the sake of brevity.) To $\cal O$($\epsilon^{\frac{1}{2}}$),
$\Psi=\epsilon^\frac{1}{2} \psi^{(1)}$ and, thus, Eq. ($\ref{ampeqn}$)
takes the form of a {\it cubic} Stuart-Landau equation:
\begin{equation}
\frac{d\Psi}{d\tau} \,=\, \epsilon \mu \Psi \,+\,
\eta \vert \Psi \vert^{2}  \Psi.
\end{equation}
\noindent
Note that $\mu(=\mu_{11})\equiv\mu_r+\imath\mu_i$ and 
$\eta(=\eta_1)\equiv\eta_r+\imath\eta_i$ 
are complex coefficients. $\Psi$ is the complex order parameter
variable whose steady state solution gives its amplitude (squared) as 
\begin{equation}
   \vert\Psi\vert^2= -\epsilon\frac{\mu_r}{\eta_r}
\label{ampsqd}
\end{equation}
\noindent
and its frequency,
 $\Omega$, (with $\Psi=\vert\Psi\vert e^{\imath\Omega\tau}$) as
\begin{equation}
{\Omega}={\epsilon}\left(\mu_i -\frac{\eta_i}{\eta_r}\mu_r\right).
\label{freq}
\end{equation}
This solution exists provided $\eta_r$ is negative since $\mu_r$ 
is  positive.  
$\eta_r$ is found to be negative over a  major part of the instability
region in the $b_0 - a$ plane, as shown in Fig. 2 (the unshaded region).
In this case, since the amplitude of the order parameter grows continuously 
in proportion to $\epsilon^{\frac{1}{2}}$ (see Eq.(\ref{ampsqd})), the 
transition is continuous ( `second order' type transition ) corresponding to 
supercritical bifurcation.  
There is a relatively small portion of the instability 
region, shown in the same figure in shades, where $\eta_r$
is found to be positive implying that the transition is discontinuous 
corresponding to subcritical bifurcation. In this regime, one has to 
go to 
quintic or even higher terms in the amplitude equation for obtaining an
expression for the order parameter. We have carried out the reductive 
perturbative method further to derive the {\it quintic} 
\begin{equation}
    \frac{d}{d\tau}\Psi=\epsilon\mu\Psi + \eta\vert\Psi\vert^{2}\Psi
  +\nu\vert\Psi\vert^{4}\Psi
\label{quintic}
\end{equation}
\noindent
as well as the {\it septic} 
\begin{equation}
    \frac{d}{d\tau}\Psi=\epsilon\mu\Psi + \eta\vert\Psi\vert^{2}\Psi
  +\nu\vert\Psi\vert^{4}\Psi + \xi\vert\Psi\vert^6\Psi
\label{septic}
\end{equation}
\noindent
amplitude equations. (Expressions for the complex coefficients 
$\nu = \nu_r + \imath\nu_i$ and $\xi = \xi_r + \imath\xi_i$  
are given in Eqs. (A3) and (A4), respectively, in Appendix A.) 
The amplitude (squared), $\vert\Psi\vert^2$, and frequency,
$\Omega$, found from the steady state solution of the quintic amplitude
equation are
\begin{equation}
\vert\Psi\vert^2=\frac{1}{2}\left(-\frac{\eta_r}{\nu_r} +
\sqrt{(\frac{\eta_r}{\nu_r})^2 -4\epsilon\frac{\mu_r}{\nu_r}}\right),
\end{equation}
\noindent
and
\begin{equation}
\Omega=\epsilon\mu_i + \eta_i\vert\Psi\vert^2 + \nu_i\vert\Psi\vert^4
\end{equation}
\noindent
while these same quantities found from the steady state solution of the
septic amplitude equation are
\begin{equation}
\vert\Psi\vert^2=\frac{1}{2}\left(-\frac{\nu_r}{\xi_r} +
\sqrt{(\frac{\nu_r}{\xi_r})^2 -4\frac{\eta_r}{\xi_r}} +
\epsilon\frac{\mu_r}{\eta_r} \right) + {\cal O}(\epsilon^2),
\end{equation}
\noindent
and
\begin{equation}
\Omega=\epsilon\mu_i + \eta_i\vert\Psi\vert^2 + \nu_i\vert\Psi\vert^4 +
\xi_i\vert\Psi\vert^6.
\end{equation}
\noindent
The ranges of validity of the quintic and septic amplitude equations enable
us to describe a large portion of the subcritical bifurcation. Fig. 2 
also shows the portion within the subcritical bifurcation where the 
dynamics of the system is supposed to be governed by the {\it quintic} 
(marked by dots) and {\it septic} (marked by open circles) amplitude
equations. The rest of the domain of subcritical bifurcation where 
{\it even} higher order amplitude equation need to be considered is marked by 
crosses on the same figure.

As can be seen from Fig. 2, there are two distinct regions of subcritical 
bifurcation located on either side of $a = 0.5$. The crossover from a 
supercritical to subcritical bifurcation is first seen for $b_0 = 0.01$ 
around $a = 0.65$ as the value of $b_0$ is reduced. The extent of 
subcritical bifurcation in the $b_0 - a$ plane increases, as $b_0$ 
decreases. The validity of the amplitude equations will be examined by 
comparing the asymptotic solution derived from them with that of the limit 
cycle solution found by numerically integrating the model equations.

\section {Comparison with Experiments and Numerical Solutions}

\subsection{Comparison with Experiments}

\noindent
To start with, consider the supercritical regime where
expressions for the amplitude and period of the limit 
cycle are relatively simple. 
Using the steady state solution of the cubic amplitude equation, 
Eqs.($\ref{ampsqd}$) and ($\ref{freq}$), the dependence of 
amplitude $\vert\Psi\vert^2$ and period, 
P ($\propto\frac{1}{\omega+\Omega}$), of the limit cycle on $a$ or $b_0$ 
can be obtained. (Here, we will only consider how the two quantities,
$|\Psi|^2$ and P, depend on $b_0$ for fixed $a$. Similar analysis can be
carried out for parameter $a$ fixing $b_0$.) The parameter 
$b_0 (=\gamma/\theta V_m)$ is a function of the applied stress 
($\sigma_a$) and temperature (T). As remarked earlier, $\gamma$ represents 
the stress and thermal activation, and at temperatures of interest, 
thermal activation can be completely ignored. As for the stress activation, 
it is a threshold process and therefore, can be taken to have a week 
dependence on stress  until a critical value of this stress  beyond  which, 
it should show a rapid increase. One such functional form for $\gamma$  
could be $exp (- \sigma_c/ \sigma)$, where $\sigma_c$ represents the 
value of the stress beyond which the function rapidly rises. Since 
stress activation occurs at large values, one expects  $\sigma_c$ to 
be  large. Using the standard expression \cite{alex} for 
$V_m(\sigma_a, T) = V_0 (\sigma_a/\sigma_0)^m exp (- E_m/kT)$ 
(with $m>1$), we get   
\begin{equation}
 b_0  \sim \gamma (\sigma) \left(\frac{\sigma_a}{\sigma_0}\right)^{-m} e^{\frac{E_m}{kT}}.
\label{rel}
\end{equation}
From this we see that $b_0$ has a decreasing dependence on stress, 
with the term $\left(\frac{\sigma_a}{\sigma_0}\right)^{-m}$ dominating 
up to $\sigma_c$ beyond which it should increase. Clearly, $b_0$ 
decreases as a function of $T$. Figure 3 shows plots of $\vert\Psi\vert^2$
and $P$ as a function of $b_0$. From the above discussion on the dependence
of $b_0$ on $\sigma$ and $T$, we see that  $\vert\Psi\vert^2$ should increase
as stress increases for the {\it major part} of stress value, since 
$\sigma_c$ is large. On the other hand, it should decrease with increase 
in temperature. Since stress 
and temperature are measurable quantities, our predictions can be compared 
with experimental results. The amplitude and period of the limit cycle are 
related, respectively, to the amount of strain jumps and the period of the 
jumps on the creep curve calculated through the Orowan equation. 
There are very few experiments in this mode of testing as mentioned 
in the introduction. The only experiment where this dependence on stress 
and temperature has been measured for a limited range is that by  
Zagoruyko {\it et. al.} (in \cite{dasi}). According to them, the amplitude of 
the strain jumps increases with stress while its period 
has a  decreasing dependence on stress. Experiments from constant
strain rate case also exhibit the same trend when the results are 
translated in terms of constant stress experiments. It is well 
known that the amplitude of the stress drops decreases with 
applied strain rate. In fact, even the numerical solution of the equations 
extended to the constant strain rate case predicts this behaviour
\cite{gava}. This implies that the amplitude of strain 
jumps should  increase as stress increases \cite{view}. [ This relation can 
be seen as follows. In constant strain rate case, the deformation rate is 
fixed and the stress developed in the sample is measured.
When the contribution to the plastic strain rate increases
due to increased dislocation motion (for whatever reasons), 
the stress has to fall in order to keep the applied strain rate constant. 
Thus, the relation between strain rate and stress is opposite.] 
Clearly, the general trend is consistent with the experimental results
for most of the value of $b_0$.
Zagorukuyko {\it et. al.} also report that the amplitude
of the strain jumps increases while its period
decreases with increase in temperature, which is  consistent with our
result. 

Fig. 4 shows plots of amplitude (squared) and period of the LC as a function 
of $b_0$ for a certain interval within the subcritical bifurcation.
Using similar analysis as above, we find that while the dependence of
the amplitude on $b_0$ is consistent with the experimental results reported 
by Zagoruyko {\it  et.al.}, the dependence of the period on $b_0$ is
predominantly inconsistent with their report.

Experiments  in constant strain rate case show that the stress drops 
are seen to arise both abruptly as well as continuously  
\cite{nonl1}. Translating this result to the constant 
stress case, it implies that the strain jumps can arise  both 
abruptly and continuously. This feature is manifest in the 
supercritical and the subcritical bifurcations seen in our calculations. 

\subsection{Comparison with Numerical Solutions}

Having derived the amplitude equation, we  now 
compare its result with the numerical solutions 
obtained via Eqs. (10-12). As stated earlier, there are two distinct
types of solutions, namely the supercritical and the subcritical solutions.
Even in the region of subcritical bifurcation, we have three types - first 
where the quintic amplitude equation works, second where the septic amplitude 
equation works and lastly, the rest of the instability domain where 
even higher order nonlinearities dominate and, as such, require even 
higher order amplitude equation. We will compare the solutions obtained 
from these with the numerically exact solutions of Eqs.(10-12).

First consider the supercritical region. Using the steady state solution of 
the cubic amplitude equation, Eqs.(\ref{ampsqd}, \ref{freq}), in
Eq.(\ref{lincomb}), leads to an analytic  expression for the limit 
cycle (LC) near bifurcation points in the domain of supercritical 
bifurcation. This is usually referred to as the {\it secular equation}. 
The derivation of the equations governing the LC  are given
in Appendix B. This solution can be compared with the numerically
exact solution obtained by integrating the system of equations, Eqs.(10-12). 
Since the region of applicability in the $b_0-a$ plane is large, we 
choose one solution for large values of $b_0$ (at the top end of 
the bifurcation diagram)  and another value of $b_0$ at the lower end.
Figures 5(a) and (b) show the plots of the secular equation 
along with their respective numerical solutions 
for two widely apart transition points within the domain of supercritical
bifurcation. As can be observed clearly, they match very well.

In a manner similar to the supercritical bifurcation, the analytic
expression for the LC for a {\it large portion of the domain of the
subcritical bifurcation} is obtained by using 
the steady state solution of the quintic or the septic amplitude equations.  
In this case, there are three distinct types of solutions: 
(a) first, where $\nu_r$ is negative and $\xi_r$ positive implying that 
only the quintic amplitude equation has a steady state solution, 
(b) second, when $\nu_r$ and $\xi_r$ are both negative implying that
both the quintic and the septic amplitude equations support steady state 
solutions, and  (c) third, when $\nu_r$ is positive and 
$\xi_r$  is negative implying that the septic amplitude equation 
supports steady state solution while the quintic does not. The resulting  
LCs are compared with those found from the numerical ones. For the case (a), 
we have picked two different bifurcation points one to the right and 
another to the left of $a = 0.5$.  Plots in Figs. 6(a) and (b) show LCs 
obtained from the secular equation and from the numerical  ones.
For the case (b), Fig. 7 shows three plots of LCs two of which are obtained
from the secular equations corresponding to the quintic and septic
amplitude equations while the third one is obtained from the numerical 
solution. It is clear from this figure that the LC derived from the 
septic amplitude equation is 
a better approximation than that derived from the quintic one. 
This shows that the `containing' role is played not only by the quintic 
but also by the septic (and even higher) term of the nonlinearity. 
We also find that the region over which such a situation is 
valid is a substantial portion of the validity of the quintic amplitude 
equation suggesting that  higher order nonlinearities are in fact 
important as was indicated by our earlier work \cite{mbga2}. For the 
case (c), we have shown in Fig. 8   comparing the LC derived from the
septic amplitude equation with that of the numerical one for a bifurcation 
point where the quintic amplitude equation {\it does not} hold. The agreement
is reasonable except around the sharp turning point.

Lastly, it is worthwhile stating that expansion of amplitude equation even 
up to septic term fails to cover the entire instability domain 
eventhough most of it is covered. This also suggests that higher order 
nonlinearities control the rest of the subcritical bifurcation domain. 
This feature is not usually  encountered in model systems.

\noindent
The above comparison  reveales the following:
(i) The expression for the LC in the supercritical domain mimics the
numerical solution very well provided $\epsilon$ is taken to be small 
enough.
(ii) In the subcritical domain also the two results generally match
well with each other. However, for values of the parameters 
where the value of $c$ is close to zero, (which also corresponds 
to small values of $b_0$), higher order nonlinearities 
could supplement the contributions arising from the lower ones in 
determining the LC. In other words, if the coefficient of the next 
higher order has also negative real part its contribution may have 
to be included in the expression for the order parameter.

\section{ Summary and Discussion}

We have carried out the reductive perturbative approach to the problem
of steps on the creep curve regarding it as a formation of ordered state
when the system is driven away from equilibrium. The dynamics of the
system is described by two coupled amplitude equations: one for a transient 
order parameter $\Psi_0$ and another for the 
complex order parameter $\Psi$ in the neighbourhood of the bifurcation 
point. The order parameter $\Psi$ 
represents both the amplitude and the phase of the limit
cycle solution when $\epsilon > 0$. Since the above
derivation is valid only in the neighbourhood of the
critical value, the expression for the order parameter
$\Psi$ is valid only for small $\epsilon$. This has been exemplified by the
quite reasonable agreements between the LC solutions found from the 
analytic expression and that from the numerical integration  
of the model. We have shown that both supercritical and subcritical
bifurcations are seen in the transitions to the instability domain.  
{\it While the major part of the phase boundary exhibits
supercritical bifurcation, subcritical bifurcation gradually dominates
as the value of $b_0$ is reduced}. The results of
earlier calculation \cite{gads2} which used the method of relaxation 
oscillation
and showed `first order' type transition is consistent with the present 
one since the values of the parameters ($a=.63$, $b_0=10^{-4}$) 
falls in the subcritical bifurcation domain.
In our more recent work \cite{mbga2} where  we first adiabatically 
eliminated one of the variables and then
applied the reductive perturbative method for the reduced model we found
both supercritical and subcritical regions. 
However, due to the fact the adiabatic elimination itself was valid 
only for small values of $b_0$, the results were found to be valid 
in a small domain. 
Further, we found that very high order nonlinearities were important. This 
result is supported by the present calculation.

The present analysis shows a feature 
which is normally not seen in model systems, namely that higher order 
terms than quintic may control the subcritical bifurcation. 
In fact, even when the quintic term plays the `containing role,'
higher order terms may also contribute significantly to the 
`role of containment'.  Thus, in deriving the expression for the LC, it may be 
necessary to find out how higher and higher order nonlinear terms in the 
expansions in the amplitude equation contribute to the asymptotic solution.

We comment here on the unusual feature of the model in the 
context of phase transitions. 
For conventional models, in the language of phase transitions, 
the `free energy' is described by an expansion in power series 
of the order parameter up to sixth power. While the `free energy' for 
`second-order' phase transitions can be described by retaining up to fourth 
power in the order parameter, its is usually sufficient to retain 
the sixth power for the description of 
`first-order' phase transitions (with the appropriate signs 
for the coefficients in the expansion). In the present case, however,
we need to go to as high as eighth power (or more) of the order parameter 
in some regions of the drive parameter to describe the corresponding `free
energy' when the transition is of `first-order' type. This feature is 
rather unusual and not found in conventional models of phase transitions.

Due to the closed form expressions 
for the order parameter, the present calculation helps
us to map the theoretically introduced parameters to the 
experimentally measured quantities. {\it For instance, in experiments one
measures, the amplitude of the strain jumps on the creep curve as well
as their frequency as a function of stress and temperature. The
dependence of the amplitude of the  strain  jumps and its frequency
on stress and temperature
can be evaluated by using the order parameter equations in the Orowan's
equation by noting that} $\vert\Psi\vert$ {\it corresponds to the
amplitude of the strain jumps while the frequency} $\Omega$ together with
$\omega$
{\it gives a measure of the frequency of the steps}. 
By properly relating the parameter $b_0$ to stress and temperature
we found that for a certain range of the instability 
the amplitude increases while the period decreases
as stress increases (at constant temperature) qualitatively 
agreeing with the reported experimental result of Zagorukuyko 
{\it et al} (see in \cite{dasi}). Also the amplitude increases while
the period decreases as temperature is increased (at constant stress)
which is again consistent with the experimental results of Zagorukuyko
{\it et. al.}
In addition we found other ranges 
within the instability showing various kinds of dependence of the 
order parameter on stress and temperature.

Lastly, the present exercise has demonstrated the complicated 
dependence of the order parameter
variable on the original modes. This will serve as a warning to those
using hand waiving arguments for declaring certain modes as fast modes
and others as slow modes in the modelling of such problems.

\subsection*{ Acknowledgement}
One of us (M.B.) would like to thank the International Program in Physical
Sciences, Uppsala University (Sweden) for offering a fellowship to study
at Indian Institute of Science. This work is partially supported by IFCPAR 
project 1108-1.


\section*{Appendix A}

In this appendix we give the expressions for the coefficients appearing
in the TDGL equation, i.e. the expressions for
$\mu$, $\eta$, $\nu$ and $\xi$ in the equation 
$\frac{d}{d\tau}\Psi=\epsilon\mu\psi +\eta\vert\Psi\vert^2\Psi
  +\nu\vert\Psi\vert^4\Psi + \xi\vert\Psi\vert^6\Psi$. Note that they are 
functions of the parameters $a$, $b_0$ and $c$. They are given below.
\setcounter{equation}{0}
\renewcommand{\theequation}{A\arabic{equation}}
\begin{equation}
\mu=\mu_{11}.
\end{equation}
\begin{equation}
\eta = \sum_{k\leq\alpha\leq j,\gamma\leq\beta} (g_{1j\alpha} h_{j\beta\gamma} + 
g_{1\alpha k}h_{k\beta\gamma}).
\end{equation}
\noindent
In Eq. (A2) and all the following equations summations over {\it latin} 
alphabets run over 1,0, and -1 while those over {\it greek} alphabets run over 
$\pm 1$. Summation of Eq. (A2) includes only those terms which satisfy the 
condition $ \alpha + \beta + \gamma = 1$.
\begin{equation}
\nu = \sum_{k \leq j} (T_{jk}^{(1,4)} + T_{jk}^{(2,3)}),
\end{equation}
\noindent
where
\begin{equation}
T_{jk}^{(1,4)} = \sum_{\zeta\leq\delta\leq\gamma\leq\beta\leq\alpha} 
(g_{1j\alpha}u_{j\beta\gamma\delta\zeta} + 
g_{1\alpha k}u_{k\beta\gamma\delta\zeta})
\end{equation}
\noindent 
and
\begin{equation}
T_{jk}^{(2,3)} = \sum_{\beta \leq \alpha,\zeta\leq\delta\leq\gamma} 
(h_{j\alpha \beta}t_{k\gamma \delta \zeta} + 
h_{k\alpha \beta} t_{j\gamma\delta\zeta}),
\end{equation}
\noindent
with the condition on the summations of Eqs. (A4) \& (A5) to be 
$\alpha + \beta +\gamma  + \delta + \zeta = 1$.
\noindent
Lastly, the septic coefficient $\xi$ is given by
\begin{equation}
\xi = \sum_{k \leq j} (T_{jk}^{(1,6)} + T_{jk}^{(2,5)} + T_{jk}^{(3,4)})
\end{equation}
\noindent
where
\begin{equation}
T_{jk}^{(1,6)} = \sum_{k\leq\alpha\leq j,\sigma\leq\rho\leq\zeta\leq\delta
\leq\gamma\leq\beta\leq\alpha}(g_{1j\alpha}w_{j\beta\gamma\delta\zeta
\rho\sigma} + g_{1\alpha k}w_{k\beta\gamma\delta\zeta\rho\sigma}),
\end{equation}
\begin{equation}
T_{jk}^{2,5} = \sum_{\beta\leq\alpha,\sigma\leq\rho\leq\zeta\leq\delta
\leq\beta\leq\alpha}g_{1jk}( h_{j\alpha\beta} v_{k\gamma\delta
\zeta\rho\sigma} + h_{k\alpha\beta}v_{j\gamma\delta\zeta\rho\sigma})
\end{equation}
and
\begin{equation}
T_{jk}^{3,4} = \sum_{\gamma\leq\beta\leq\alpha,\sigma\leq\rho\leq\zeta
\leq\delta}g_{1jk}( t_{j\alpha \beta \gamma} 
u_{k\delta\zeta\rho\sigma} + t_{k\alpha\beta\gamma} 
u_{j\delta\zeta\rho\sigma})
\end{equation}
\noindent
with the constraint $\alpha + \beta + \gamma + \delta + \zeta + \rho + 
\sigma = 1$ imposed on the summations of Eqs. (A9), (A10) \& (A11).
\noindent
Expressions for $\mu_{jk}$, $g_{jkl}$, $h_{jkl}$, $t_{jklm}$, $u_{jklmn}$, 
$v_{jklmnp}$ and $w_{jklmnpq}$ are given in Appendix C.

\section*{Appendix B}
\noindent

\setcounter{equation}{0}
\renewcommand{\theequation}{B\arabic{equation}}

\noindent
In this appendix, (approximate) asymptotic solution for
$\vec{R}$ describing the limit cycle will be derived using the
steady state solution of the particular TDGL equation.

\noindent
For the instability region where cubic TDGL equation holds,
expansion of $\Psi$ upto $\psi^{(2)}_j$ will be involved in
determining $\vec{R}$  so that 
\begin{equation}
\vec{R} = \epsilon \psi^{(2)}_0 e^{\lambda_o\tau} \vec{r}_0 + \left[
(\epsilon^{\frac {1} {2}} \psi^{(1)} + \epsilon \psi^{(2)} )
e^{iw\tau} \vec{r} + c.c \right]
\end{equation}

\noindent
Note that the asymptotic solution does not contain $\psi^{(1)}_0
e^{\lambda_0\tau}$ term since, $\lambda_0$ being negative, it
decays in time. To $\cal O$($\epsilon^{\frac{1}{2}}$)
\begin{equation}
\Psi \,=\, \epsilon^{\frac{1}{2}} \psi^{(1)} \, \Rightarrow \, \psi^{(1)}
\,=\, \epsilon^{-\frac{1}{2}} \Psi.
\end{equation}
\noindent
Using Eqs. (B2) and (32) in Eq. (B1) will enable us to derive the components
of $\vec{R}$ which are given by: 
\begin{equation}
X = A_x\vert\Psi\vert cos(\Omega_c\tau + \theta_x)
+ B_{1x}\vert\Psi\vert^2 cos(2\Omega_c\tau + \phi_x) + B_{2x}\vert\Psi\vert^2+
B_{3x}\vert\Psi\vert^2 cos(2\Omega_c\tau +\varphi_x),
\end{equation}
\begin{equation}
Y = A_y\vert\Psi\vert cos(\Omega_c\tau + \theta_y)
+ B_{1y}\vert\Psi\vert^2 cos(2\Omega_c\tau + \phi_y) + B_{2y}\vert\Psi\vert^2+
B_{3y}\vert\Psi\vert^2 cos(2\Omega_c\tau +\varphi_y)
\end{equation}
\noindent
and
\begin{equation}
Z = A_z\vert\Psi\vert cos(\Omega_c\tau + \theta_z)
+ B_{1z}\vert\Psi\vert^2 cos(2\Omega_c\tau + \phi_z) + B_{2z}\vert\Psi\vert^2+
B_{3z}\vert\Psi\vert^2 cos(2\Omega_c\tau +\varphi_z).
\end{equation}
\noindent
where
\begin{equation}
\Omega_c=\Omega + \omega,
\end{equation}
\begin{equation}
A_d=2\vert r_{1d}\vert,
\end{equation}
\begin{equation}
\theta_d=sin^{-1}\left(\frac{Im(r_{1d})}{\vert r_{1d}\vert}\right),
\end{equation}
\begin{equation}
B_{1d}=2\vert h_{111}r_{1d}\vert,
\end{equation}
\begin{equation}
\phi_d=sin^{-1}\left(\frac{Im(h_{111}r_{1d})}{\vert h_{111}r_{1d}\vert}\right),
\end{equation}
\begin{equation}
B_{2d}=2\vert h_{11-1}r_{1d}\vert,
\end{equation}
\begin{equation}
B_{3d}=2\vert h_{1-1-1}r_{1d}\vert,
\end{equation}
\noindent
and
\begin{equation}
\varphi_d=sin^{-1}\left(\frac{-Im(h_{1-1-1}r_{1d})}{\vert h_{1-1-1}r_{1d}\vert}\right).
\end{equation}
\noindent
Here the subscript $d$ denotes $x$, $y$ or $z$ while $Im$ denotes imaginary
part of the concerned argument. Expressions for $h_{jkl}$ and $r_{jd}$
are given in Appendix C.

\noindent
For the instability region where quintic (or septic) TDGL equation
holds, expansion of $\Psi$ upto $\psi^{(4)}_j$ (or upto $\psi^{(6)}_j$)
will be required in determining $\vec{R}$ so that
\begin{equation}
\vec{R} = \sum^N_{n=2} \epsilon^{n/2} \psi^{(n)}_0
e^{\lambda_o\tau} \vec{r}_0 + \sum^N_{n=1}[ \epsilon^{\frac{n}{2}} 
\psi^{(n)} e^{i\omega\tau} \vec{r} + c.c ]
\end{equation}
\noindent
with N=4 (quintic case), N= 6 (septic case). The same procedure as done 
for the cubic case follows for the quintic as well as the septic cases.

\section*{Appendix C}
\noindent
\setcounter{equation}{0}
\renewcommand{\theequation}{C\arabic{equation}}
\noindent
In this appendix we first give the expressions for $h_{kl,j}$, $t_{klm,j}$,
$u_{klmn,j}$, $v_{klmnp,j}$ and $w_{klmnpq,j}$ that appear as coefficients 
in the determination of $\psi^{(2)}_j$, $\psi^{(3)}_j$, $\psi^{(4)}_j$, 
$\psi^{(5)}$ and $\psi^{(6)}$, respectively, such that
\begin{equation}
\psi^{(2)}_j e^{\lambda_{j} \tau} \,=\, \sum _{l \leq k} h_{jkl}
\psi^{(1)}_{k} \psi^{(1)}_{l} e^{(\lambda_{k} \,+\, \lambda_{l}) \tau},
\end{equation}
\begin{equation}
\psi^{(3)}_{j} e^{\lambda_{j} \tau}=\sum _{m \leq l\leq k} t_{jklm}
\psi^{(1)}_k \psi^{(1)}_l\psi^{(1)}_m e^{(\lambda_k+\lambda_l+\lambda_m) \tau},
\end{equation} 
\begin{equation}
\psi^{(4)}_{j} e^{\lambda_{j} \tau}=\sum _{n\leq m\leq l\leq k}u_{jklmn}
\psi^{(1)}_k \psi^{(1)}_l\psi^{(1)}_m \psi^{(1)}_n 
e^{(\lambda_k+\lambda_l+\lambda_m+\lambda_n) \tau},
\end{equation} 
\begin{equation}
\psi^{(5)}_j e^{\lambda_j\tau} = \sum_{p \leq n \leq m \leq l \leq k} 
v_{jklmnp}\psi^{(1)}_k \psi^{(1)}_l \psi^{(1)}_m \psi^{(1)}_n \psi^{(1)}_p
e^{(\lambda_k + \lambda_l +\lambda_m + \lambda_n + \lambda_p)\tau},  
\end{equation}
\noindent
and
\begin{equation}
\psi^{(6)}_j e^{\lambda_j\tau}=\sum_ {q\leq p\leq n\leq m\leq l\leq k}       
w_{jklmnpq}\psi^{(1)}_k\psi^{(1)}_l\psi^{(1)}_m\psi^{(1)}_n
\psi^{(1)}_p\psi^{(1)}_q
e^{(\lambda_k + \lambda_l +\lambda_m + \lambda_n + \lambda_p+\lambda_q)\tau}.
\end{equation}
\noindent
We give only those coefficients that contribute to the asymptotic solution. 
Note that summation over {\it latin} alphabets take values 1,0 and -1 while
summation over {\it greek} alphabets take values $\pm 1$.
\begin{equation} 
h_{j\alpha \beta} = \frac {g_{j\alpha \beta}}
{\lambda_{\alpha} + \lambda_{\beta} - \lambda_{j}}.
\end{equation}

\begin{equation}
t_{j\alpha\beta\gamma} =\frac{1}{C_{j\alpha\beta\gamma}}
\sum_{l\leq\alpha\leq k} (g_{j\alpha l} h_{l\beta \gamma} +
g_{jk \alpha} h_{k\beta \gamma}) 
\end{equation}
\noindent
provided $\alpha + \beta + \gamma \neq j$. Otherwise, 
$t_{j\alpha\beta\gamma}=0$. $C_{j\alpha\beta\gamma}=
\lambda_{\alpha} + \lambda_{\beta} + \lambda_{\gamma} - \lambda_j$. 

\begin{equation}
u_{j\alpha \beta \gamma \delta} = \frac{1}{C_{j\alpha\beta\gamma\delta}}
\sum_{k,l}\left[(g_{j\alpha l} t_{l\beta \gamma \delta} +
g_{jk \alpha} t_{k\beta \gamma \delta}) +  g_{jk l}
(h_{k\alpha \beta} h_{l\gamma \delta})\right]
\end{equation}
\noindent
where $C_{j\alpha\beta\gamma\delta}= \frac{1}{ \lambda_{\alpha} +
\lambda_{\beta} + \lambda_{\gamma} - \lambda_j}$.
 
\begin{equation}
v_{j\alpha \beta \gamma \delta \zeta} = \frac{1}
{C_{j\alpha\beta\gamma\delta\zeta}}
\sum_{k,l}\left[(g_{j\alpha l} u_{l\beta \gamma \delta\zeta} +
g_{jk \alpha} u_{k\beta \gamma \delta\zeta}) +  g_{jk l}
(h_{k\alpha \beta} t_{l\gamma \delta\zeta} + h_{l\alpha \beta}
t_{k\gamma \delta \zeta})\right]
\end{equation}
\noindent
provided $\alpha + \beta + \gamma + \delta + \zeta \neq j$. Otherwise, 
$v_{j\alpha\beta\gamma\delta\zeta}=0$. The expression
$C_{j\alpha\beta\gamma\delta\zeta}= \frac{1}{\lambda_{\alpha} +
\lambda_{\beta} + \lambda_{\gamma} + \lambda_{\delta} + \lambda_{\zeta} 
- \lambda_j}$.

\begin{equation}
w_{j\alpha\beta\gamma\delta\zeta\rho} = \frac{1}
{C_{j\alpha\beta\gamma\delta\zeta\rho}}
\sum_{k,l}\left[Q + g_{jkl}
(h_{k\alpha \beta} u_{l\gamma \delta\zeta\rho} + h_{l\alpha \beta}
u_{k\gamma \delta \zeta \rho}) + g_{jk l} t_{k\alpha
\beta \gamma} t_{l\delta\zeta\rho}\right]
\end{equation}
\noindent
where $Q=(g_{j\alpha l} v_{l\beta\gamma\delta\zeta\rho} +
g_{jk \alpha} v_{k\beta\gamma\delta\zeta\rho})$ and
$C_{j\alpha\beta\gamma\delta\zeta\rho}=\frac{1}{\lambda_{\alpha} +
\lambda_{\beta} + \lambda_{\gamma} + \lambda_{\delta} + \lambda_{\zeta}
+ \lambda_{\rho} - \lambda_j}$.

\noindent
Next we give the expressions for $\mu_{jk}$, $\eta_j$ and $g_{jkl}$: 
\begin{equation}
\mu_{jk}=\frac{c_0s_{jz}(-r_{kx} + r_{kz})}{\vec{s}^T_{j}\vec{r}_j},
\end{equation}
\begin{equation}
\eta_j = \sum_{l\leq\alpha\leq k,\gamma\leq\beta} (g_{jk\alpha} h_{k\beta\gamma} + 
g_{j\alpha l}h_{l\beta\gamma}),
\end{equation}
\begin{equation}
for \, k \neq l,
g_{jkl} =2f_{xxj} r_{kx}r_{lx} + f_{xyj}(r_{kx}r_{ly} + r_{ky}r_{lx}), 
\end{equation}
\noindent
while
\begin{equation}
g_{jkk} =f_{xxj} r^2_{kx} + f_{xyj}r_{kx}r_{ky},
\end{equation}
\noindent
where 
\begin{equation}
f_{xxj}=\frac{b_0(-s_{jx}+b_0 s_{jy})}{\vec{s}^T_{j}\vec{r}_j},
\end{equation}
\begin{equation}
f_{xyj} = \frac{-(s_{jx} + b_0 s_{jy})}{\vec{s}^T_j\vec{r}_j},
\end{equation}
\begin{equation}
\vec{r}^T_{j} =(r_{jx} \, r_{jy}\, r_{jz}),
\end{equation}
\begin{equation}
\vec{s}^T_{j} =(s_{jx} \, s_{jy}\, s_{jz}),
\end{equation}
\begin{equation}
\vec{s}^T_j\vec{r}_j = s_{jx}r_{jx} + s_{jy}r_{jy} + s_{jz}r_{jz},
\end{equation}
\begin{equation}
r_{jx}=\chi(c_{0}+\lambda_{j}),
\end{equation}
\begin{equation}
r_{jy}=-(c_{0} + \lambda_{j})(\alpha + \lambda_{j}),
\end{equation}
\begin{equation}
r_{jz}=c_0\chi,
\end{equation}
\begin{equation}
s_{jx}=-(c_{0} + \lambda_{j})(b_0\delta + \lambda_{j}),
\end{equation}
\begin{equation}
s_{jy}=r_{jx},
\end{equation}
\noindent
and
\begin{equation}
s_{jz}=ab_0\chi.
\end{equation}

\newpage

\begin{figure}
\caption{  The instability region determined by all the three
independent parameters $a$, $b_0$ and $c$ of the model. It is 
bounded by the three surfaces: $c_0$-surface (shown by a series of
curved lines, $c=0$ plane (shown by a series of straight lines), and 
$b_0=0$ plane. (Note that the $c_0$ surface is determined from Eq. (23).)}
\label{fig1}
\end{figure}
\begin{figure}
\caption{ Plot of the bifurcation diagram in the $a$ - $b_0$
plane. The instability region is bounded by the parabolic-shaped 
curve and the $b_0=0$ line. The unshaded(shaded) region exhibits 
super-(sub-)critical bifurcation. The shaded region marked by dots 
(open circles) shows the portion within the sub-critical 
bifurcation where quintic (septic) amplitude equation is supposed to hold.}
\label{fig2}
\end{figure}
\begin{figure}
\caption{  Plots of $|\Psi|^2$ and $P$ versus $b_0$ within 
the supercrtical domain when $a=0.45$ and $\epsilon=0.01$.}
\label{fig3}
\end{figure}
\begin{figure}

\caption{  Plots of $|\Psi|^2$ and $P$ versus $b_0$ within 
the subcritical domain when $a=0.39$ and $\epsilon=10^{-4}$.}
\label{fig4}
\end{figure}
\begin{figure}

\caption{  Plots of the limit cycle solutions (in X-Z plane) 
obtained from the secular equation (solid line) and that 
obtained from numerical integration of the model (Eqs. 10-12) 
(marked with dots) when (a) $a=0.521$, $b_0=0.0202$ and $\epsilon=0.1$
and when (b) $a=0.6$, $b_0=0.01$ and $\epsilon=0.001$. Both bifurcation
points lie within the supercritical bifurcation domain.}
\label{fig5a,b}
\end{figure}
\begin{figure}

\caption{  Plots of the limit cycle solutions (in X-Z plane) 
obtained from the secular equation (solid line) and that 
obtained from numerical integration of the model (Eqs. 10-12) 
(marked with dots) when (a) $a=0.647$, $b_0=0.009$ and $\epsilon=0.01$
and when (b) $a=0.38$, $b_0=0.004$ and $\epsilon=0.0001$. Both bifurcation
points lie within the subcritical bifurcation domain where the quintic
amplitude equation holds (i.e. $\nu_r < 0$ and $\xi_r > 0$).}
\label{fig6a,b}
\end{figure}
\begin{figure}

\caption{  Plots of the limit cycle solution (in X-Z plane) 
obtained by using the secular equation derived from the quintic amplitude 
equation (outer solid line), from the septic amplitude equation 
(inner solid line) and that obtained from numerical integration of 
the model (Eqs. 10-12) (marked with dots) when  $a=0.375$, $b_0=0.004$ 
and $\epsilon=0.0001$. The bifurcation point lies within the subcritical 
bifurcation domain where {\it both} $\nu_r$ and $\xi_r$ are negative.}
\label{fig7}
\end{figure}
\begin{figure}

\caption{ Plots of the limit cycle solution (in X-Z plane) 
obtained from the secular equation (solid line) 
and that obtained from numerical integration of 
the model (Eqs. 10-12) (marked with dots) when  $a=0.37$, $b_0=0.001$ 
and $\epsilon=0.0001$. The bifurcation point lies within the subcritical 
bifurcation domain where  $\nu_r > 0$ and $\xi_r < 0$.}
\label{fig8}
\end{figure}

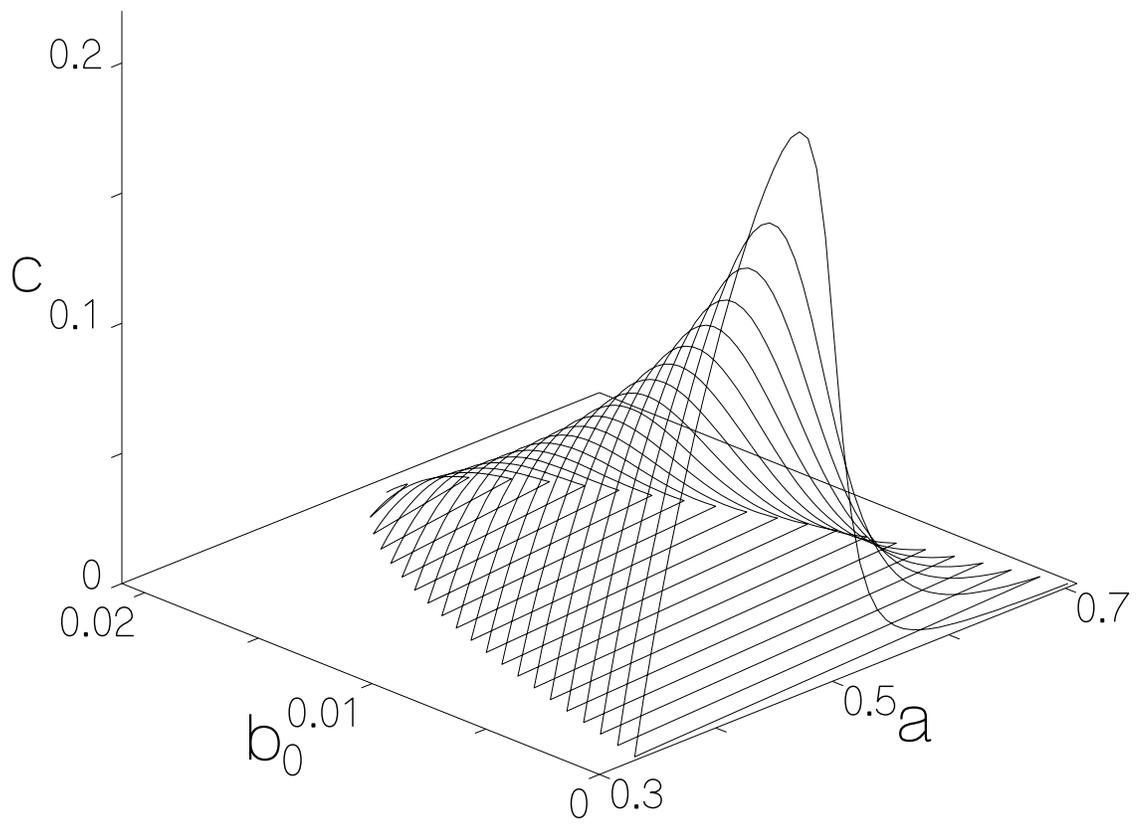
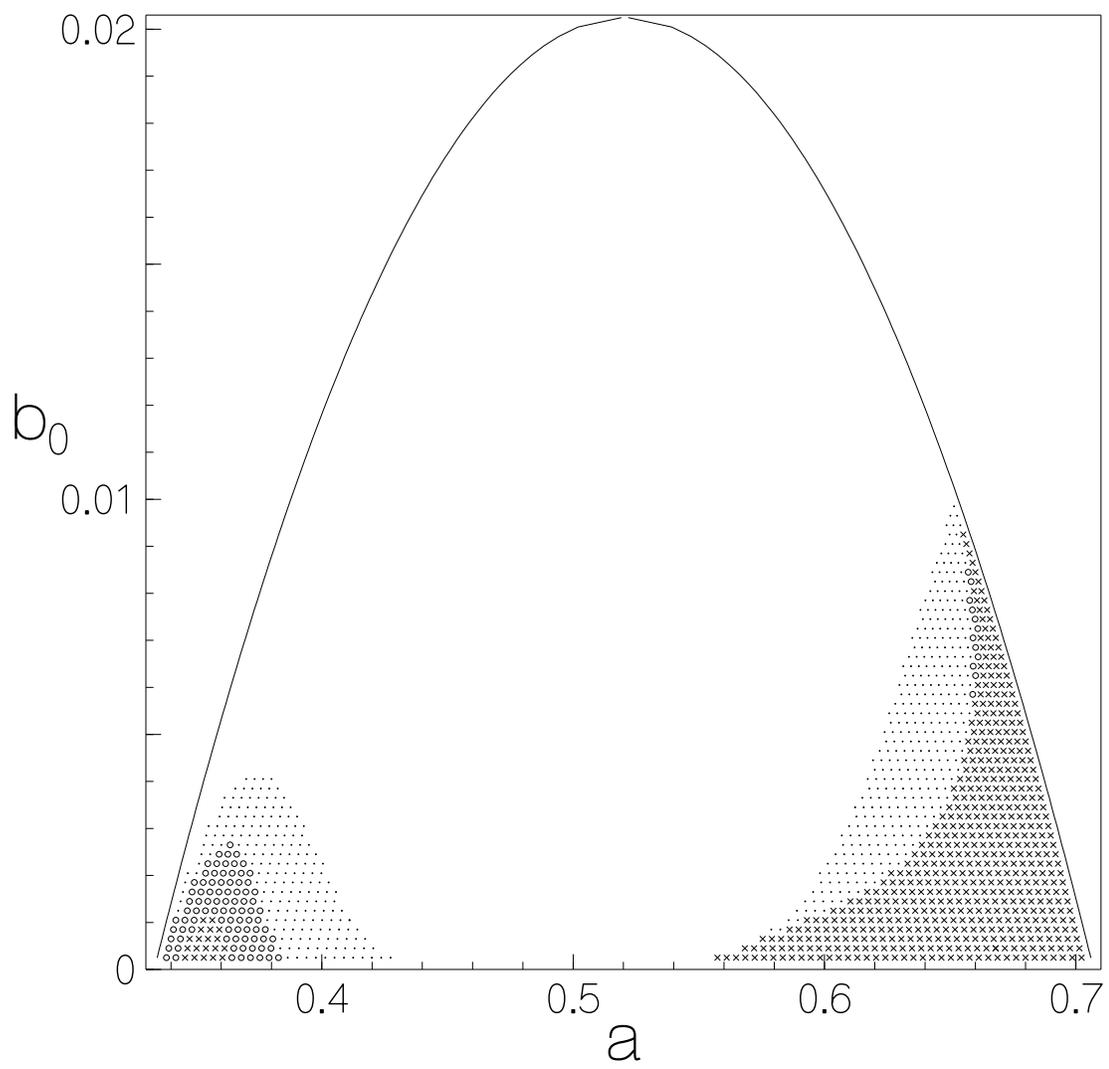
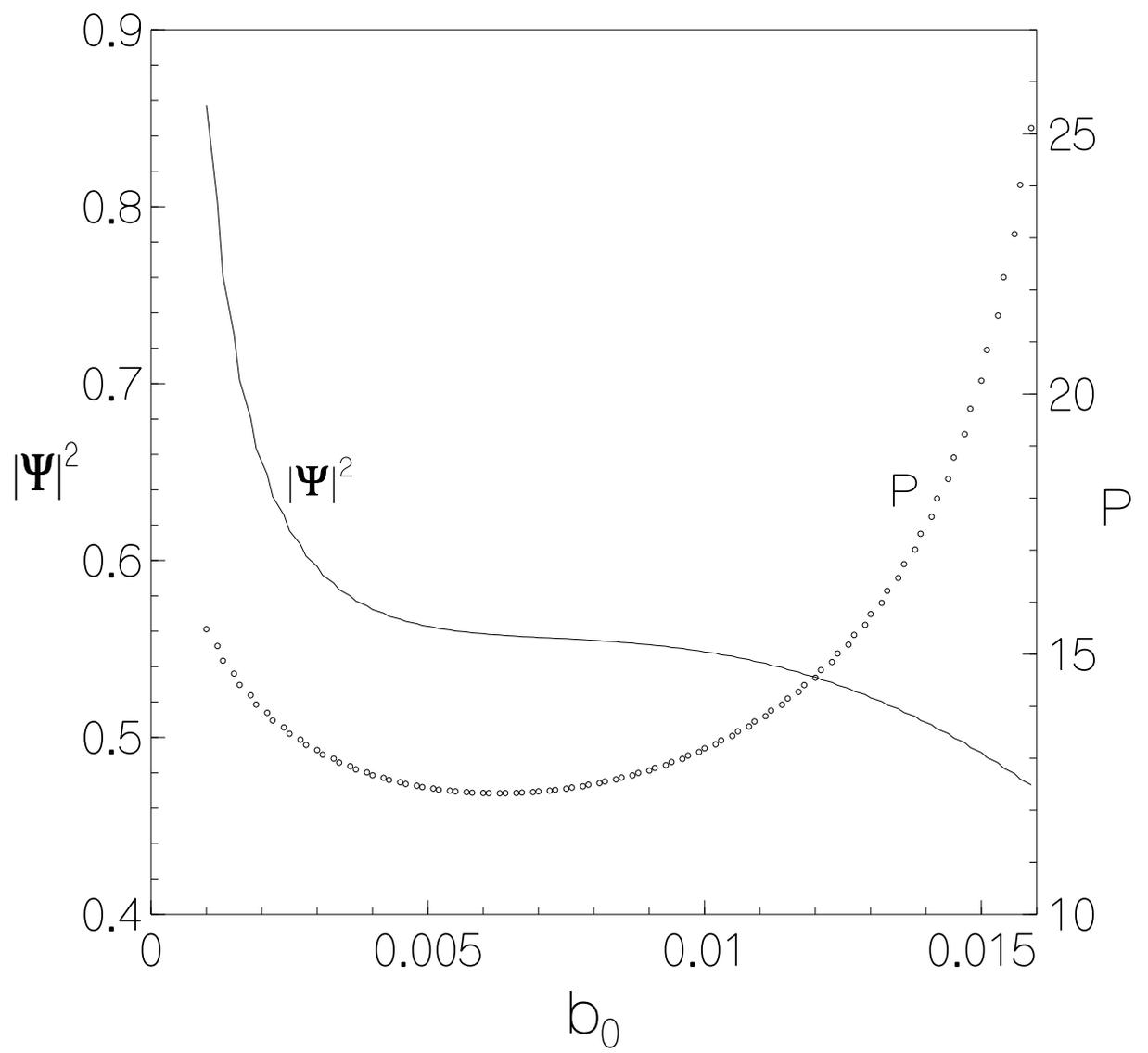
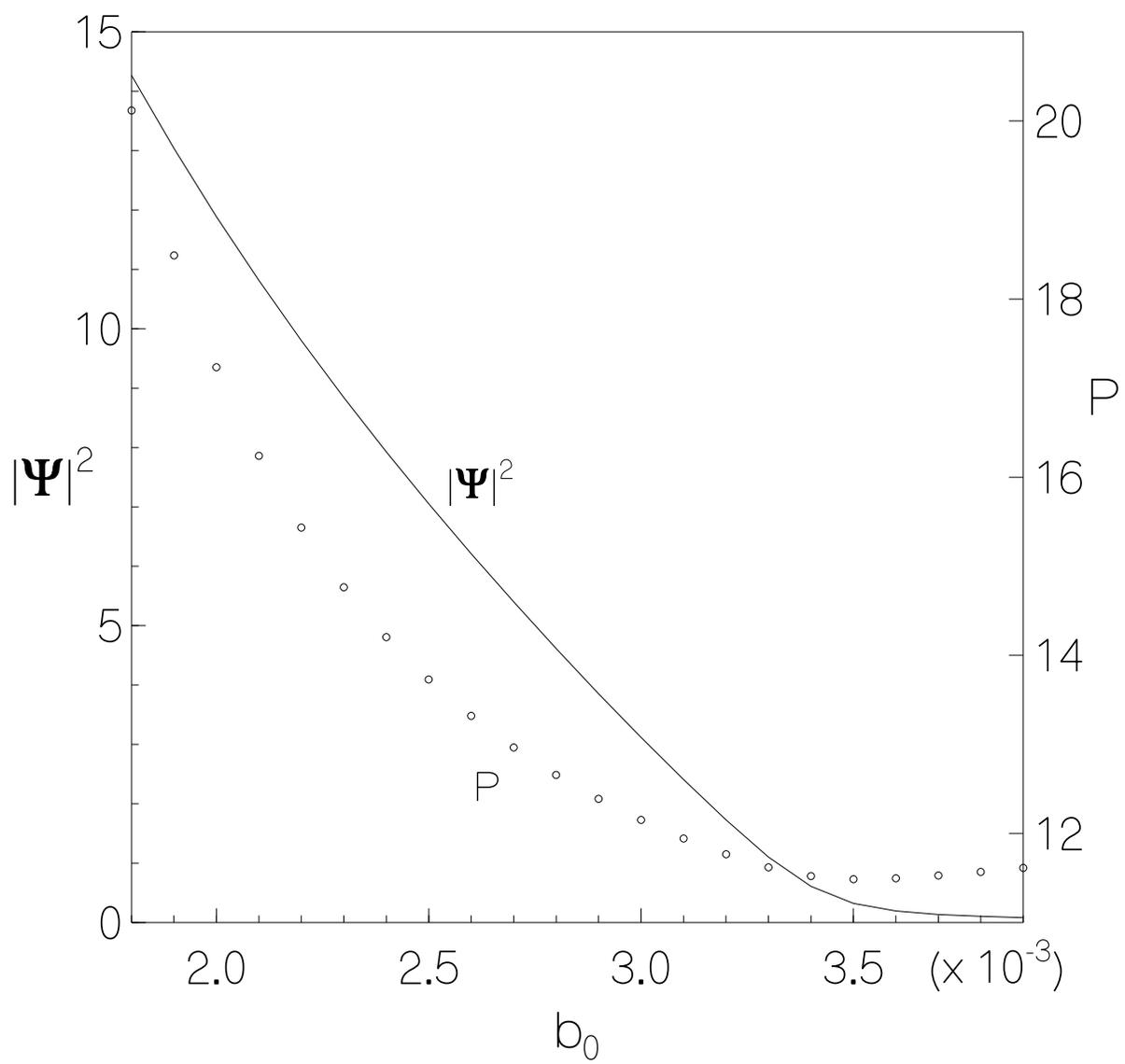
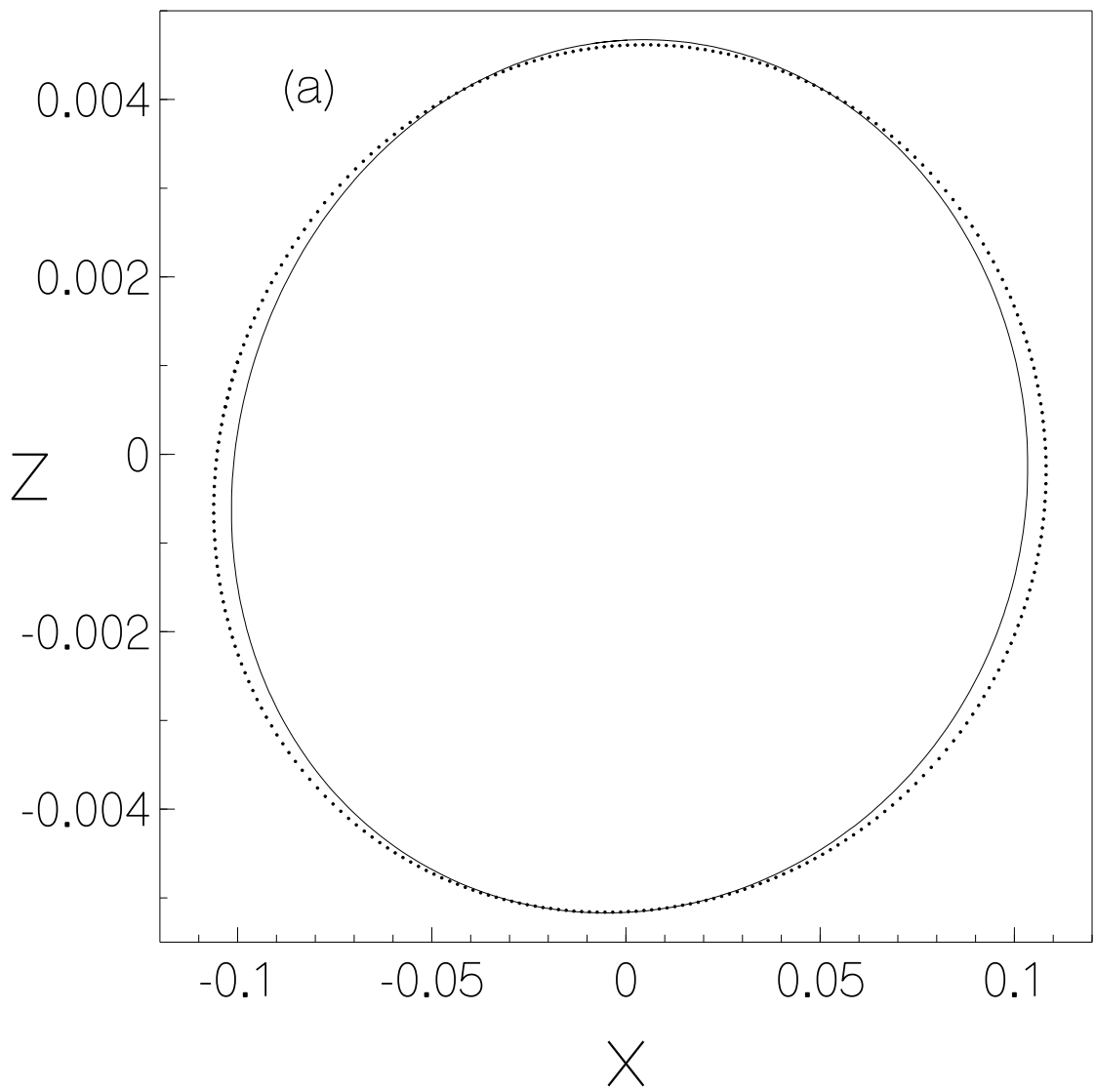
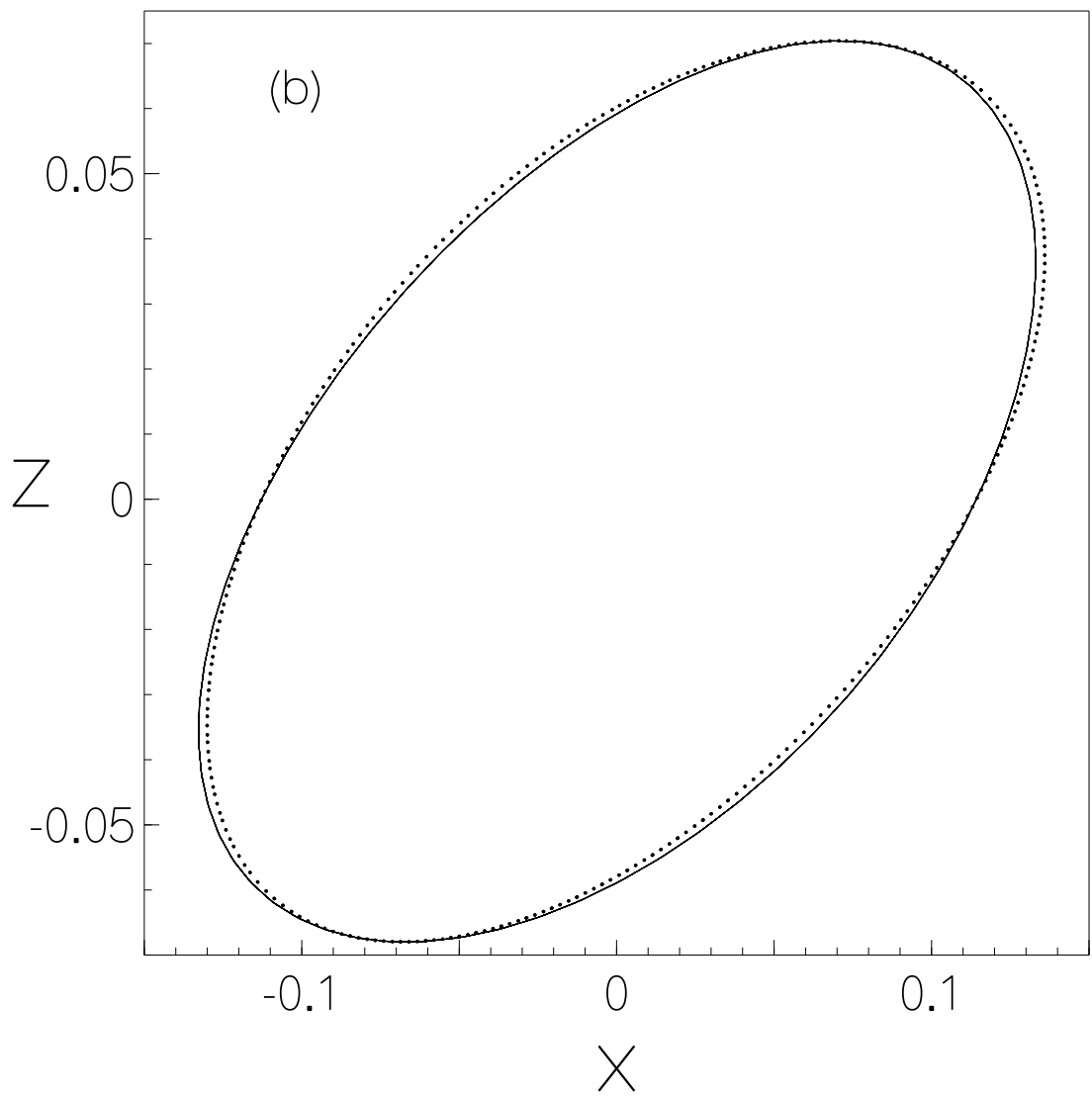
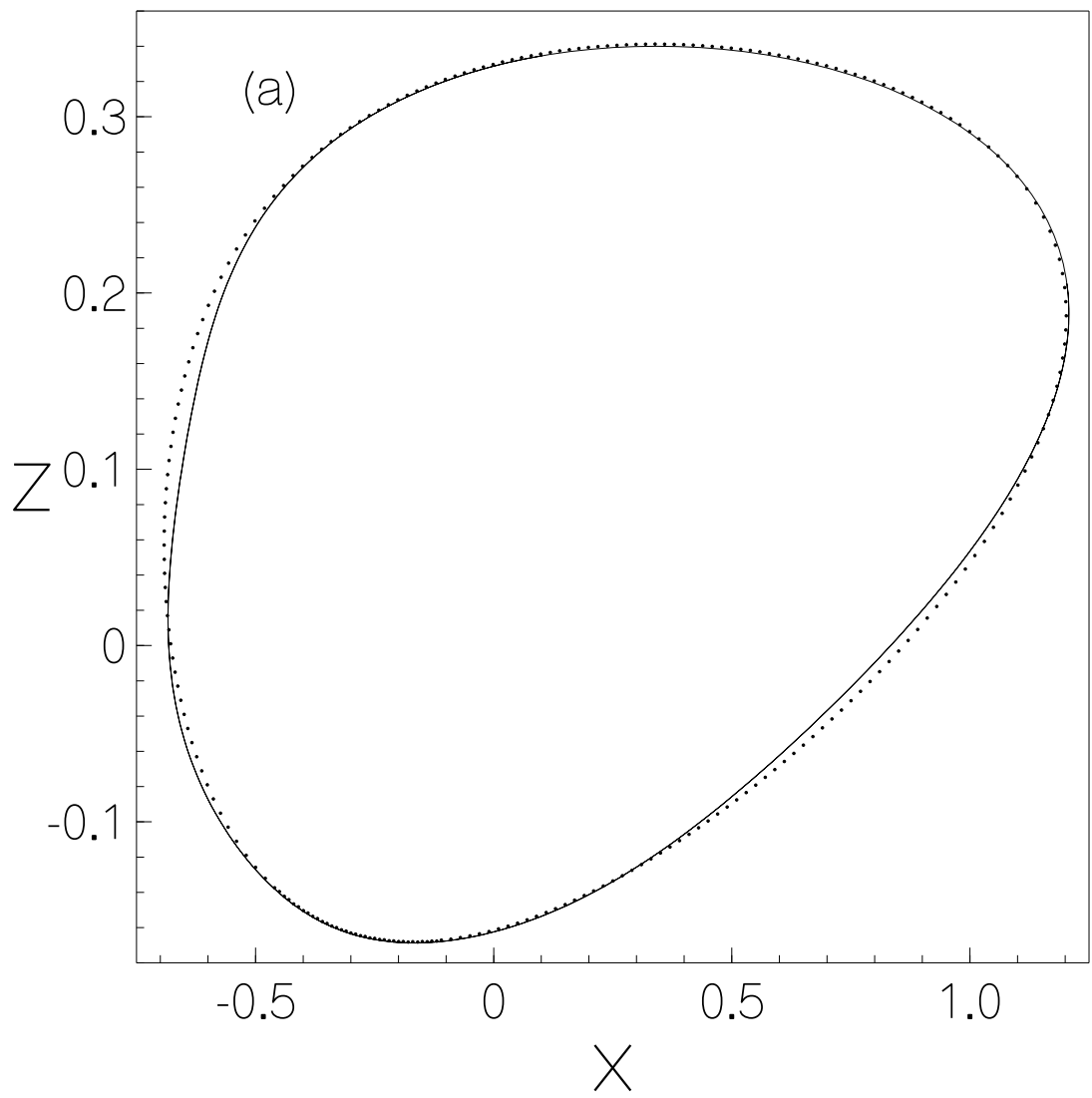
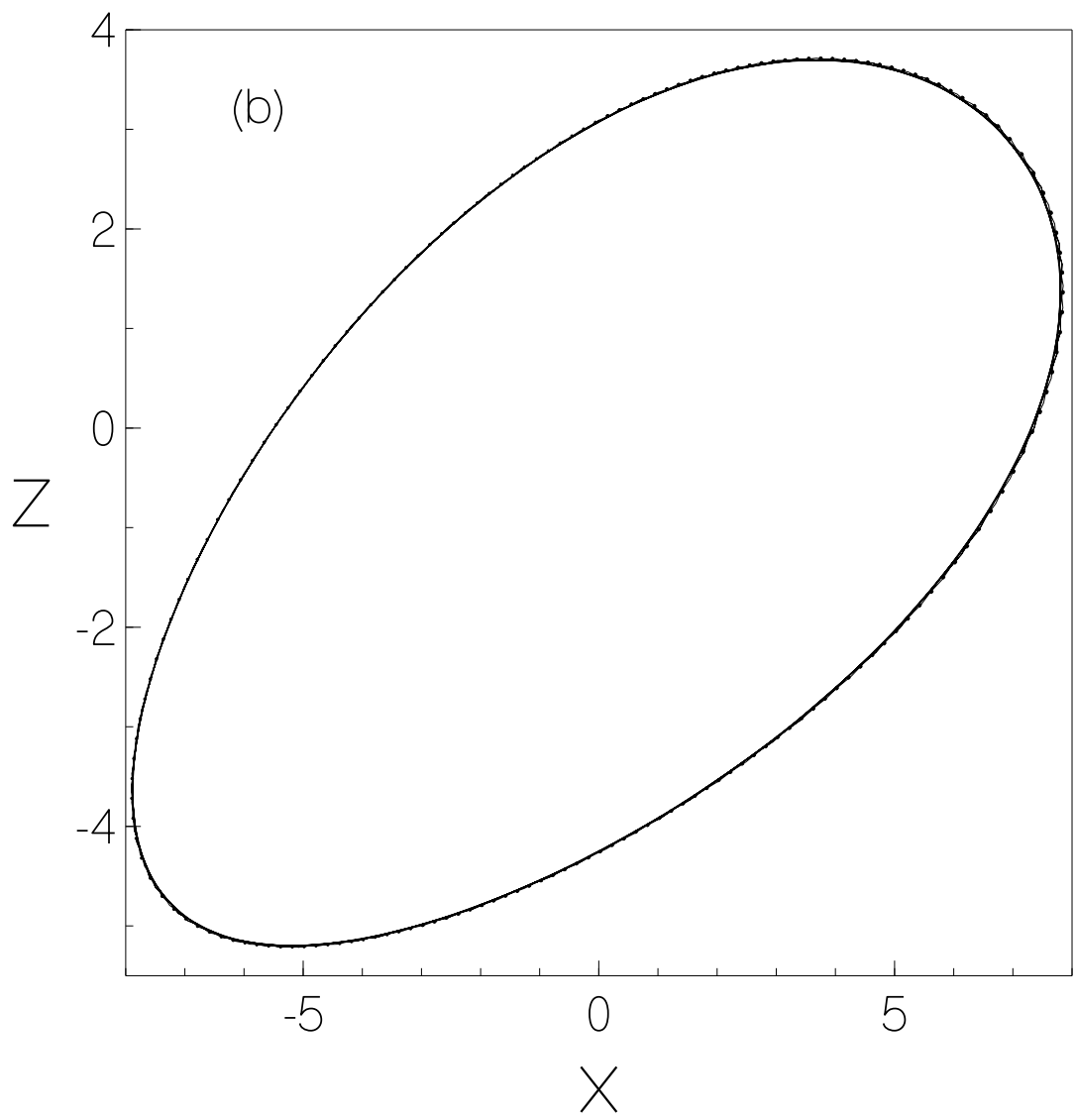
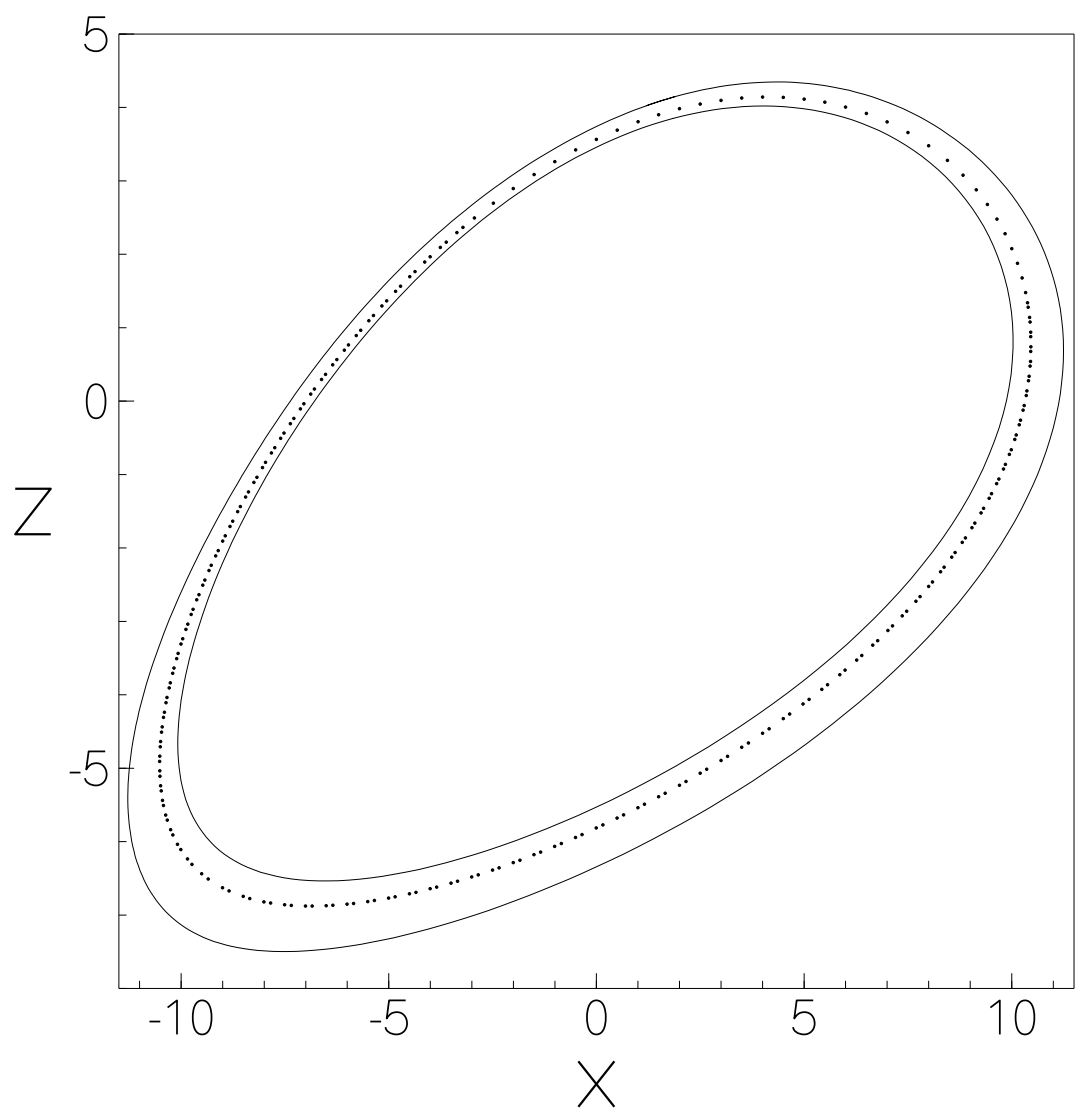
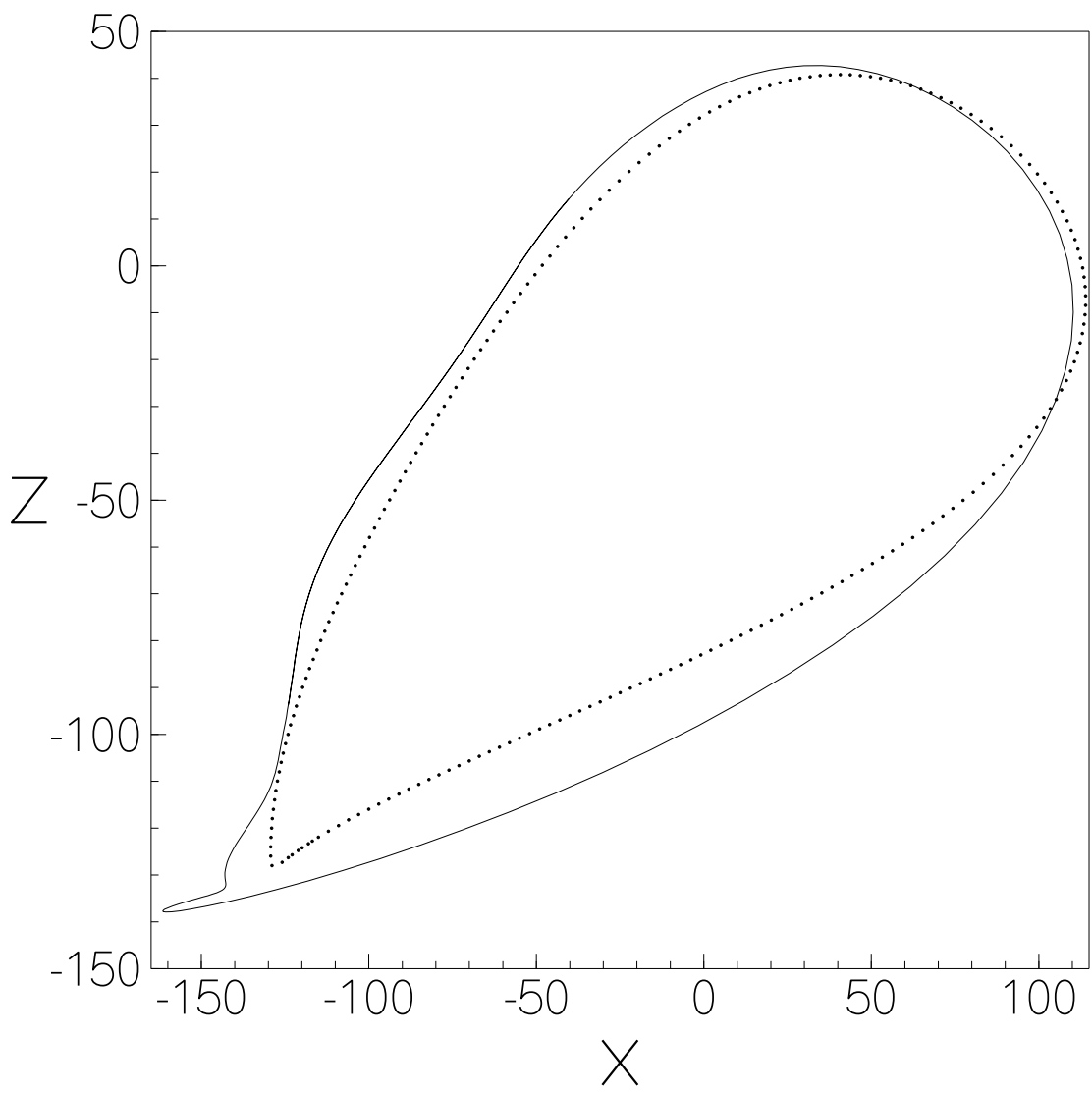
\end{document}